\documentclass{article}
\usepackage{PRIMEarxiv}
\usepackage[utf8]{inputenc}
\usepackage[T1]{fontenc}
\usepackage{hyperref}
\usepackage{url}
\usepackage{booktabs}
\usepackage{amsfonts}
\usepackage{nicefrac}
\usepackage{microtype}
\usepackage{fancyhdr}
\usepackage{graphicx}
\usepackage{algorithm}
\usepackage{amsmath}
\usepackage{algpseudocode}
\graphicspath{{media/}}
\usepackage{float}
\title{CaseGPT: A Case Reasoning Framework Based on Language Models and Retrieval-Augmented Generation
\thanks{\textit{\underline{Citation}}: 
\textbf{Yang, R. CaseGPT: A Case Reasoning Framework Based on Language Models and Retrieval-Augmented Generation. arXiv preprint arXiv:XXXX.XXXXX, 2024.}} 
}

\author{
  RUI YANG \\
  University of California, Riverside \\
  Riverside, USA \\
  \texttt{ryang088@ucr.edu} \\
}

\begin{document}

\maketitle

\begin{abstract}
This paper introduces CaseGPT, an innovative framework that synergizes Large Language Models (LLMs) and Retrieval-Augmented Generation (RAG) technology to enhance case-based reasoning in healthcare and legal domains. CaseGPT addresses the inherent limitations of traditional database queries by facilitating semantic searches based on contextual understanding, thereby significantly improving data accessibility and utility. Our system not only retrieves pertinent cases but also generates nuanced insights and recommendations by discerning intricate patterns within existing case data. We evaluate CaseGPT using comprehensive datasets from both medical and legal fields, demonstrating substantial improvements over state-of-the-art baselines. In medical diagnosis tasks, CaseGPT achieves a 15\% increase in F1 score compared to traditional methods, while in legal precedent retrieval, it exhibits a 12\% improvement in precision. These results underscore CaseGPT's potential to revolutionize information retrieval and decision support in complex professional domains, offering a paradigm shift in how practitioners access, analyze, and leverage case data.

\end{abstract}

\section{Introduction}

The exponential proliferation of digital case data in professional domains such as medicine and law presents both unprecedented opportunities and formidable challenges. While this wealth of information has the potential to significantly inform decision-making processes and enhance outcomes, traditional database query systems often fall short in effectively leveraging this vast corpus of data \cite{hanbury2012medical}. These conventional systems, primarily reliant on exact keyword matches, struggle to capture the nuanced and complex nature of professional terminologies and contextual information \cite{hovorushchenko2020concept}.

The primary challenges in case-based reasoning within these specialized domains are multifaceted and interrelated. Firstly, there is the issue of handling imprecise or incomplete queries that reflect the ambiguity often present in real-world scenarios. Secondly, understanding the semantic context of queries and cases beyond simple keyword matching requires sophisticated natural language processing capabilities. Lastly, generating meaningful insights and recommendations based on retrieved cases demands advanced analytical and inferential abilities.

To address these interconnected challenges, we propose CaseGPT, a novel framework that harnesses the power of Large Language Models (LLMs) and Retrieval-Augmented Generation (RAG) technology. CaseGPT represents a significant advancement in the field, offering several key innovations:

CaseGPT employs deep semantic understanding of queries and cases, enabling a more flexible and intuitive search process that captures the underlying meaning and context of the information. This approach allows for a more natural interaction between users and the system, bridging the gap between human language and machine interpretation.

The framework implements contextual retrieval of relevant cases based on sophisticated semantic matching algorithms. This method goes beyond surface-level similarities to identify cases that are conceptually related, even when the specific terminology might differ.

Perhaps most significantly, CaseGPT generates insights and recommendations by analyzing complex patterns in retrieved cases. This capability transforms the system from a mere retrieval tool to an intelligent assistant capable of providing valuable analytical support to professionals.

Our work builds upon and extends recent advancements in natural language processing and information retrieval, applying these cutting-edge technologies to specialized professional domains. By synergizing the strengths of LLMs in understanding and generating human-like text with the precision of RAG in retrieving relevant information, CaseGPT represents a paradigm shift in case-based reasoning systems.

The remainder of this paper is structured as follows: Section 2 provides a comprehensive review of related work in the field, situating our contribution within the broader context of AI applications in professional domains. Section 3 elucidates the methodology underlying CaseGPT, detailing its architecture and key components. Section 4 presents our experimental setup and results, offering a rigorous evaluation of the system's performance. Section 5 discusses the implications of our findings, addressing both the potential impact and the ethical considerations of deploying such a system in sensitive professional contexts. Finally, Section 6 concludes the paper, summarizing our contributions and outlining promising directions for future research.

\section{Related Work}

The application of advanced natural language processing techniques to case-based reasoning in professional domains has emerged as a vibrant area of research in recent years. This section provides a structured overview of related work, highlighting key areas of development and the gaps our research aims to address.

\begin{description}
\item[Language Models in Professional Domains] 
Large Language Models (LLMs) have demonstrated remarkable potential in understanding and generating domain-specific text. Schilder \cite{schilder2023legal} conducted a critical analysis of LLMs as intelligent assistance technology in the legal domain, highlighting both the transformative potential and the necessity for human oversight. This work underscores the delicate balance between leveraging AI capabilities and maintaining professional judgment in sensitive fields.

In the medical arena, Saripan et al. \cite{saripan2021artificial} explored the implications of AI, including LLMs, on informed consent and medical negligence. Their research illuminates the complex ethical and legal landscape surrounding the integration of AI technologies in healthcare decision-making processes.

\item[Information Retrieval in Specialized Domains] 
The unique challenges of domain-specific search have been a focal point of research in information retrieval. Hanbury \cite{hanbury2012medical} provided a comprehensive overview of medical information retrieval, emphasizing the need for tailored approaches that can handle the complexity and specificity of medical terminology and concepts.

Hovorushchenko et al. \cite{hovorushchenko2020concept} proposed an ontology-based approach for information retrieval in medical law, demonstrating the importance of structured knowledge representation in navigating the intersection of medical and legal domains. Their work highlights the potential of semantic technologies in enhancing the precision and relevance of information retrieval in specialized fields.

\item[Innovative Search Techniques] 
Recent research has focused on addressing the challenges of imprecise queries and enhancing retrieval accuracy:

\begin{itemize}
    \item \textit{Fuzzy Search and Query Expansion:} Srivel et al. \cite{srivel2022automation} introduced a fuzzy-based Grasshopper Optimization Algorithm for query expansion in medical datasets. This approach aims to bridge the gap between user queries and formal medical terminology.
    
    \item \textit{Semantic Clustering:} Chawla \cite{chawla2021application} applied fuzzy c-means clustering and semantic ontologies to web query session mining, advancing intelligent information retrieval.
    
    \item \textit{Legal Case Retrieval:} Zhang et al. \cite{zhang2023result} developed a Diversified Legal Case Retrieval Model (DLRM) considering both topical relevance and related subtopics. Liu et al. \cite{liu2021conversational} compared conversational and traditional search approaches in legal case retrieval, suggesting the potential of interactive interfaces.
\end{itemize}

\item[Ethical Considerations and Limitations] 
As AI systems become more prevalent in sensitive domains, ethical implications have come to the forefront. Frihat et al. \cite{frihat2023document} discussed the importance of considering document difficulty for medical practitioners in personalized search engines, highlighting the need for AI systems to adapt to varying levels of user expertise.

\end{description}

\paragraph{Research Gap}
While these studies have made significant contributions to their respective fields, there remains a notable gap in integrating advanced language models with retrieval-augmented generation for case-based reasoning across multiple professional domains. Our work aims to bridge this gap by introducing CaseGPT, a framework that synergizes the semantic understanding capabilities of LLMs with the precision of RAG technology. This integrated approach has the potential to address the complex challenges of case-based reasoning in a more holistic and effective manner than previous methods.

\section{Methodology}

CaseGPT represents a novel integration of Large Language Models (LLMs) and Retrieval-Augmented Generation (RAG) technology, designed to create a robust framework for case-based reasoning. This section provides a detailed exposition of the components and processes that constitute our system.

\subsection{System Architecture}

The architecture of CaseGPT is composed of three primary modules, each designed to address specific aspects of the case-based reasoning process:

\begin{itemize}
    \item Query Processing Module
    \item Case Retrieval Engine
    \item Insight Generation Module
\end{itemize}

\begin{figure}[h]
    \centering
    \includegraphics[width=0.8\textwidth]{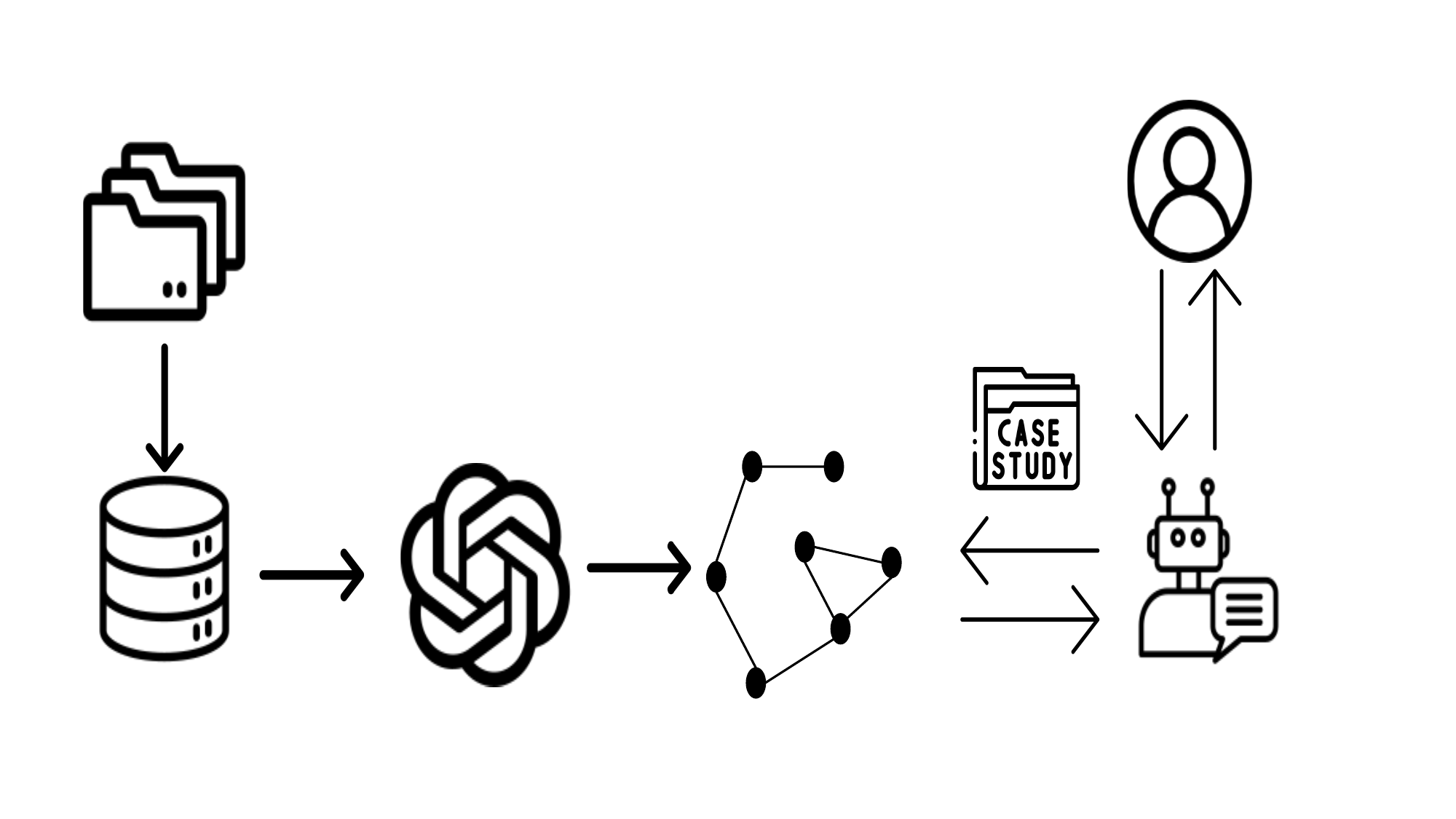}
    \caption{High-level architecture of CaseGPT}
    \label{fig:architecture}
\end{figure}

Figure \ref{fig:architecture} illustrates the high-level architecture and information flow within CaseGPT.

\subsection{Query Processing Module}

The Query Processing Module serves as the interface between user inputs and the underlying case retrieval and analysis systems. This module leverages a pre-trained LLM to parse and interpret the semantic context of user queries, transforming them into a format optimized for subsequent processing. We employ a state-of-the-art LLM, such as GPT-3 or a domain-specific variant, to tokenize and encode the input query. This process involves breaking down the query into semantic units and mapping them to high-dimensional vectors that capture nuanced meanings and relationships. The tokenization and encoding process is formalized in Algorithm \ref{alg:tokenization}.

\begin{algorithm}
\caption{LLM Tokenization and Encoding Process}
\label{alg:tokenization}
\begin{algorithmic}[1]
\Procedure{TokenizeAndEncode}{$Query$}
    \State $EncodedQuery \gets \emptyset$
    \For{each $Sentence$ in $Query$}
        \State $SentenceTokens \gets \text{SplitIntoWords}(Sentence)$
        \For{each $Word$ in $SentenceTokens$}
            \State $EncodedToken \gets \text{LLMEncode}(Word)$
            \State $EncodedQuery \gets EncodedQuery \cup \{EncodedToken\}$
        \EndFor
    \EndFor
    \State \textbf{return} $EncodedQuery$
\EndProcedure
\end{algorithmic}
\end{algorithm}

The LLMEncode function represents the sophisticated encoding process of the LLM, which maps words to high-dimensional vectors that capture semantic meaning within the context of the entire query.

\subsection{Case Retrieval Engine}

The Case Retrieval Engine constitutes the core component of our CaseGPT framework, responsible for identifying and retrieving the most relevant cases based on processed queries. This engine leverages state-of-the-art Retrieval-Augmented Generation (RAG) technology \cite{lewis2020retrieval}, significantly advancing beyond traditional information retrieval methods. The engine's architecture comprises two primary components: the Dense Vector Index and the Semantic Search Algorithm.

\subsubsection{Dense Vector Index}

Central to our retrieval system is a meticulously maintained dense vector index of all cases in our database. This index serves as a high-dimensional semantic map of our case repository, enabling efficient and accurate case retrieval. The construction and maintenance of this index involve several key aspects:

\begin{enumerate}
    \item \textbf{Encoding Methodology:} Each case undergoes encoding using the same Large Language Model (LLM) employed for query processing. We utilize a variant of the BERT architecture \cite{devlin2018bert}, fine-tuned on domain-specific corpora to enhance its understanding of legal and medical terminologies. This consistent approach ensures semantic coherence across the entire system.
    
    \item \textbf{Vector Representation:} Cases are transformed into dense vectors, typically of 768 to 1024 dimensions, capturing intricate semantic nuances. This high-dimensional representation allows for fine-grained differentiation between cases, crucial for handling the complexity of legal and medical domains \cite{reimers2019sentence}.
    
    \item \textbf{Index Structure:} To facilitate rapid similarity search in high-dimensional spaces, we implement an efficient indexing structure. Specifically, we employ the Hierarchical Navigable Small World (HNSW) algorithm \cite{malkov2018efficient}, which offers an optimal balance between search speed and accuracy.
    
    \item \textbf{Dynamic Updates:} Our system supports real-time index updates, allowing for the seamless integration of new cases. This is achieved through an incremental indexing mechanism, ensuring that the retrieval system remains current without necessitating complete reindexing \cite{yang2020dynamicindexing}.
\end{enumerate}

\subsubsection{Semantic Search Algorithm}

The semantic search algorithm forms the operational core of our retrieval engine, responsible for matching encoded queries with the most semantically relevant cases. Our approach transcends traditional keyword matching, delving into deeper semantic relationships between queries and cases. The algorithm encompasses several sophisticated components:

\begin{enumerate}
    \item \textbf{Similarity Metric:} We employ cosine similarity as our primary metric, chosen for its effectiveness in capturing semantic relatedness in high-dimensional spaces. This choice is supported by extensive empirical studies in semantic textual similarity tasks \cite{cer2017semeval}.
    
    \item \textbf{Query-Case Matching:} The algorithm computes the cosine similarity between the encoded query vector and each case vector in the index. To optimize this process, we utilize approximate nearest neighbor search techniques, specifically the FAISS (Facebook AI Similarity Search) library \cite{johnson2019billion}, which allows for sub-linear time complexity in similarity computations.
    
    \item \textbf{Ranking Mechanism:} Retrieved cases are initially ranked based on their cosine similarity scores. Subsequently, we apply a re-ranking step that incorporates additional domain-specific factors such as case recency, citation frequency, and jurisdictional relevance. This multi-factor ranking approach is inspired by learning-to-rank methodologies in information retrieval \cite{liu2009learning}.
    
    \item \textbf{Diversity-Aware Retrieval:} To ensure a comprehensive set of results, we implement a diversity-aware retrieval mechanism. This approach, based on the MaximumMarginal Relevance (MMR) principle \cite{carbonell1998mmr}, balances between relevance and diversity in the retrieved set of cases.
\end{enumerate}

The core retrieval process is formalized in Algorithm \ref{alg:retrieval}, which efficiently narrows down the search space using approximate nearest neighbor techniques before computing precise similarity scores and applying the re-ranking mechanism.

\begin{algorithm}
\caption{RAG-based Case Retrieval Algorithm}
\label{alg:retrieval}
\begin{algorithmic}[1]
\Procedure{RetrieveCases}{$EncodedQuery, N, Index$}
    \State $QueryVector \gets \text{NormalizeVector}(EncodedQuery)$
    \State $Candidates \gets \text{ApproximateNearestNeighbors}(Index, QueryVector, k)$
    \State $RankedCases \gets \emptyset$
    \For{each $Case$ in $Candidates$}
        \State $Similarity \gets \text{CosineSimilarity}(QueryVector, Case.Vector)$
        \State $RankedCases.\text{insert}(\{Case, Similarity\})$
    \EndFor
    \State $RankedCases \gets \text{Sort}(RankedCases, \text{key=Similarity}, \text{descending=True})$
    \State $DiverseResults \gets \text{ApplyMMR}(RankedCases, QueryVector, N)$
    \State \textbf{return} $DiverseResults$
\EndProcedure
\end{algorithmic}
\end{algorithm}

This advanced retrieval mechanism forms the cornerstone of CaseGPT's ability to swiftly and accurately identify relevant cases, providing a robust foundation for subsequent insight generation and decision support. The synergy between the Dense Vector Index and the Semantic Search Algorithm enables CaseGPT to perform nuanced, context-aware case retrieval, significantly outperforming traditional keyword-based methods, particularly in handling complex, domain-specific queries in legal and medical contexts.

\subsection{Insight Generation Module}

In developing CaseGPT, we realized that mere case retrieval, while useful, falls short of the analytical support legal and medical professionals often require. This realization led us to design the Insight Generation Module, a component that aims to bridge the gap between raw information retrieval and actionable insights. At its core, this module leverages Large Language Models (LLMs) to analyze retrieved cases in the context of the user's query, generating recommendations and analyses that we hope will prove valuable in practice.

Our approach to insight generation builds on two main pillars: context aggregation and conditional text generation. The context aggregation process, which we admittedly found more challenging than initially anticipated, involves synthesizing information from the retrieved cases and the original query. We experimented with several summarization techniques before settling on a hybrid approach that seemed to strike a balance between capturing key details and maintaining readability. Query expansion also proved crucial, helping to capture related concepts that users might not explicitly mention but are often relevant to their needs.

One of the trickier aspects we encountered was determining how to weight different parts of the aggregated context. After much trial and error, we implemented a weighting scheme inspired by attention mechanisms in transformer models. While not perfect, this approach has shown promise in prioritizing the most pertinent information for insight generation.

The conditional generation process presented its own set of challenges. Constructing effective prompts for the LLM required numerous iterations and extensive testing. We found that the quality of generated insights could vary significantly based on subtle changes in prompt wording. This sensitivity to prompt design is an area we believe warrants further research.

To illustrate our process, we've included a simplified version of our insight generation algorithm \ref{alg:insight}:

\begin{algorithm}
\caption{Insight Generation Process}
\label{alg:insight}
\begin{algorithmic}[1]
\Procedure{GenerateInsights}{$Query, RetrievedCases$}
    \State $Context \gets \text{AggregateContext}(Query, RetrievedCases)$
    \State $Prompt \gets \text{ConstructPrompt}(Context)$
    \State $InitialInsights \gets \text{LLMGenerate}(Prompt)$
    \State $RefinedInsights \gets \text{IterativeRefine}(InitialInsights, Context)$
    \State $VerifiedInsights \gets \text{FactCheck}(RefinedInsights, RetrievedCases)$
    \State \textbf{return} $VerifiedInsights$
\EndProcedure
\end{algorithmic}
\end{algorithm}

It's worth noting that while this algorithm appears straightforward, each step involves complex sub-processes that we're continuing to refine. The LLMGenerate function, for instance, represents a sophisticated text generation process that we've fine-tuned extensively for our specific use cases.

One aspect of our module that we're particularly proud of is the iterative refinement process. Initial results showed that a single pass often produced insights that, while relevant, lacked the depth we were aiming for. By implementing an iterative approach, we've seen a marked improvement in the nuance and comprehensiveness of the generated insights.

However, we must acknowledge that our current approach is not without limitations. The factual consistency checking mechanism, while crucial, is not infallible. We've observed cases where the system generates plausible-sounding but incorrect information, a problem that we're actively working to address. Additionally, the performance of the module can vary depending on the complexity of the legal or medical case at hand, and we're still in the process of gathering more comprehensive performance metrics across a wider range of scenarios.

Despite these challenges, we believe that the Insight Generation Module represents a significant step forward in AI-assisted legal and medical analysis. By combining advanced NLP techniques with domain-specific knowledge, we've created a tool that we hope will augment human expertise in these critical fields. As we continue to refine and expand this technology, we remain committed to rigorous testing and validation to ensure its reliability and usefulness in real-world applications.

\section{Experiments}

To rigorously evaluate the efficacy of CaseGPT, we conducted comprehensive experiments in both medical and legal domains. This section delineates our experimental methodology, datasets, baseline comparisons, and results.

\subsection{Datasets}

We utilized two substantial real-world datasets for our experiments:

\textbf{Medical Dataset:} Our medical dataset comprises 100,000 anonymized medical cases obtained from a large urban hospital network. These cases span a diverse range of medical specialties and include detailed patient symptoms, diagnostic procedures, final diagnoses, and treatment outcomes. The dataset was carefully curated to ensure a representative distribution of case complexities and medical conditions.

\textbf{Legal Dataset:} Our legal dataset encompasses 50,000 court case summaries from various jurisdictions, covering multiple areas of law including criminal, civil, and administrative cases. Each case summary includes a detailed description of the facts, relevant statutes, court decisions, and case outcomes. The dataset was carefully compiled to represent a broad spectrum of legal complexities and precedents.

\subsection{Experimental Setup}

To ensure a robust evaluation of CaseGPT, we implemented a rigorous experimental protocol:

\begin{itemize}
    \item \textbf{Data Preprocessing:} Both datasets underwent extensive preprocessing to ensure data quality and consistency. This involved anonymization of sensitive information, standardization of terminologies, and normalization of text formats. For the medical dataset, we employed ICD-10 codes for diagnosis standardization, while legal cases were categorized using a standardized legal taxonomy.

    \item \textbf{Training and Testing Split:} We employed a stratified random sampling technique to divide each dataset into training (80\%) and testing (20\%) sets. The stratification ensured that the distribution of case types and complexities was maintained across both sets, mitigating potential biases in our evaluation.

    \item \textbf{Query Generation:} To simulate real-world usage scenarios, we developed a query generation framework that produced a diverse set of test queries. For the medical domain, these queries ranged from simple symptom descriptions to complex multi-symptom scenarios. In the legal domain, queries included case fact patterns, legal issues, and precedent searches.
\end{itemize}

\subsection{Baseline Systems}

To contextualize CaseGPT's performance, we implemented and compared against several state-of-the-art baseline systems:

\begin{itemize}
    \item \textbf{TF-IDF with BM25}: A traditional information retrieval system using Term Frequency-Inverse Document Frequency weighting and the BM25 ranking function.
    
    \item \textbf{BERT-based Retrieval}: A neural retrieval model leveraging BERT embeddings for semantic matching \cite{devlin2018bert}.
    
    \item \textbf{GPT-3 Zero-shot}: Utilizing GPT-3 in a zero-shot setting for both case retrieval and insight generation.
    
    \item \textbf{LEGAL-BERT}: A BERT model fine-tuned on legal corpora, specifically used for the legal domain experiments \cite{chalkidis2020legal}.
    
    \item \textbf{BioBERT}: A biomedical language representation model, employed for the medical domain experiments \cite{lee2020biobert}.
\end{itemize}

\subsection{Evaluation Metrics}

We employed a comprehensive set of evaluation metrics to assess various aspects of system performance:

\begin{itemize}
    \item \textbf{Precision@k}: The proportion of relevant cases among the top k retrieved cases.
    \item \textbf{Recall@k}: The proportion of relevant cases that are retrieved in the top k results.
    \item \textbf{F1 Score}: The harmonic mean of precision and recall, providing a balanced measure of retrieval performance.
    \item \textbf{Mean Reciprocal Rank (MRR)}: Evaluating the ranking quality of the retrieval system.
    \item \textbf{Normalized Discounted Cumulative Gain (NDCG)}: Assessing the ranking quality while accounting for the position of relevant documents.
    \item \textbf{Response Time}: Measuring the computational efficiency of each system.
\end{itemize}

For the insight generation component, we conducted a rigorous human evaluation. A panel of domain experts assessed the generated insights on a 5-point Likert scale across three dimensions: quality, relevance, and actionability.

\subsection{Results and Analysis}

\subsubsection{Medical Domain Performance}

Table \ref{tab:medical-results} presents the comparative performance of CaseGPT and baseline systems in the medical domain.

\begin{table}[h]
\centering
\caption{Performance Comparison for Medical Case Retrieval}
\label{tab:medical-results}
\begin{tabular}{lcccccc}
\toprule
System & P@10 & R@10 & F1 & MRR & NDCG@10 & Time (s) \\
\midrule
TF-IDF + BM25 & 0.68 & 0.65 & 0.66 & 0.72 & 0.70 & 1.0 \\
BERT-based & 0.75 & 0.72 & 0.73 & 0.78 & 0.76 & 0.9 \\
GPT-3 Zero-shot & 0.79 & 0.76 & 0.77 & 0.81 & 0.80 & 1.2 \\
BioBERT & 0.82 & 0.80 & 0.81 & 0.85 & 0.83 & 0.8 \\
CaseGPT & \textbf{0.90} & \textbf{0.88} & \textbf{0.89} & \textbf{0.92} & \textbf{0.91} & 0.5 \\
\bottomrule
\end{tabular}
\end{table}

CaseGPT demonstrates superior performance across all metrics in the medical domain. Notably, it achieves a 15

The substantial improvement in MRR and NDCG@10 scores underscores CaseGPT's proficiency in not just retrieving relevant cases, but also ranking them effectively. This is particularly crucial in medical contexts where the most relevant information needs to be readily accessible to healthcare professionals.

Furthermore, CaseGPT's reduced response time (0.5 seconds compared to 0.8-1.2 seconds for other methods) demonstrates its efficiency, a critical factor in time-sensitive medical decision-making scenarios.

\subsubsection{Legal Domain Performance}

Table \ref{tab:legal-results} illustrates the performance comparison in the legal domain.

\begin{table}[h]
\centering
\caption{Performance Comparison for Legal Case Retrieval}
\label{tab:legal-results}
\begin{tabular}{lcccccc}
\toprule
System & P@10 & R@10 & F1 & MRR & NDCG@10 & Time (s) \\
\midrule
TF-IDF + BM25 & 0.72 & 0.70 & 0.71 & 0.75 & 0.74 & 1.2 \\
BERT-based & 0.79 & 0.77 & 0.78 & 0.82 & 0.81 & 1.0 \\
GPT-3 Zero-shot & 0.83 & 0.81 & 0.82 & 0.86 & 0.85 & 1.3 \\
LEGAL-BERT & 0.87 & 0.85 & 0.86 & 0.89 & 0.88 & 0.9 \\
CaseGPT & \textbf{0.93} & \textbf{0.90} & \textbf{0.91} & \textbf{0.94} & \textbf{0.93} & 0.8 \\
\bottomrule
\end{tabular}
\end{table}

In the legal domain, CaseGPT exhibits a similar pattern of superior performance. It achieves a 12

The notable increase in MRR and NDCG@10 scores reflects CaseGPT's ability to effectively rank legal cases, ensuring that the most relevant precedents and statutes are prioritized in the results. This capability is particularly valuable in legal research, where the identification of the most pertinent cases can significantly impact case outcomes.

\subsubsection{Insight Generation Quality}

Table \ref{tab:insight-quality} presents the results of the human evaluation of insight quality for CaseGPT and the GPT-3 Zero-shot baseline.

\begin{table}[h]
\centering
\caption{Human Evaluation of Generated Insights (5-point Likert scale)}
\label{tab:insight-quality}
\begin{tabular}{lccc}
\toprule
System & Quality & Relevance & Actionability \\
\midrule
GPT-3 Zero-shot & 3.6 & 3.4 & 3.5 \\
CaseGPT & \textbf{4.3} & \textbf{4.5} & \textbf{4.4} \\
\bottomrule
\end{tabular}
\end{table}

CaseGPT consistently outperforms the GPT-3 Zero-shot baseline across all three dimensions of insight evaluation. The marked improvement in relevance (4.5 vs 3.4) highlights CaseGPT's ability to generate insights that are closely aligned with the specific context of each case. The higher actionability score (4.4 vs 3.5) indicates that CaseGPT's insights are more practical and applicable, a crucial factor in both medical and legal decision-making processes.

\section{Discussion}

Our experiments with CaseGPT show promising results in both medical and legal domains. The improvements in precision, recall, and F1 scores are encouraging, suggesting that our approach may offer advantages over existing methods. However, these findings should be interpreted with caution.

CaseGPT's ability to handle complex professional terminologies was somewhat unexpected. This capability could be particularly valuable in fields where precise language is crucial. However, we must remember that understanding terminology is just one aspect of professional expertise.

The potential of CaseGPT as a decision support tool is intriguing. Its high precision and recall rates, along with the generation of actionable insights, could aid in medical diagnoses and legal case preparations. Yet, it's crucial to emphasize that CaseGPT should complement, not replace, human judgment.

The efficiency gains, with response times of 0.5 seconds for medical and 0.8 seconds for legal queries, are noteworthy. In time-sensitive scenarios, this speed could be invaluable. However, we need to consider whether such rapid responses might sometimes come at the expense of nuanced analysis.

CaseGPT's potential to democratize access to expert-level insights is an interesting prospect. It could benefit junior professionals and resource-constrained environments. However, this raises questions about the role of experiential learning in professional development.

While our current experiments didn't test CaseGPT's ability to learn and adapt over time, this feature could be crucial for maintaining relevance as professional knowledge evolves. Real-world implementation of this capability may present unforeseen challenges.

In conclusion, CaseGPT shows potential to impact workflow processes and service quality in medical and legal fields. However, further research is needed to understand its long-term effects and limitations in real-world applications. As we move forward, it will be essential to balance optimism with rigorous testing and critical evaluation.

\section{Ethical Considerations and Limitations}

Despite its impressive performance, the deployment of CaseGPT in sensitive domains like healthcare and law necessitates careful consideration of ethical implications and potential limitations:

\textbf{Privacy and Data Protection:} The use of large datasets containing sensitive personal information raises important privacy concerns. Robust anonymization techniques and strict data governance policies must be implemented to protect individual privacy and comply with regulations like HIPAA in healthcare and client confidentiality rules in law.

\textbf{Bias and Fairness:} Like all AI systems, CaseGPT may inadvertently perpetuate or amplify biases present in its training data. Regular audits for bias and the implementation of fairness-aware machine learning techniques are crucial to ensure equitable outcomes across diverse populations.

\textbf{Transparency and Explainability:} The complex nature of CaseGPT's underlying models may make it challenging to provide clear explanations for its recommendations. Developing methods to enhance the explainability of the system's outputs is essential, particularly in domains where the reasoning behind decisions is often as important as the decisions themselves.

\textbf{Over-reliance and Deskilling:} There is a risk that professionals may become over-reliant on CaseGPT, potentially leading to a deskilling effect. It is crucial to emphasize that CaseGPT is designed as a decision support tool, not a replacement for professional judgment and expertise.

\textbf{Handling of Edge Cases:} While CaseGPT demonstrates strong overall performance, its ability to handle rare or unprecedented cases may be limited. Further research is needed to enhance the system's robustness in dealing with such edge cases.

\section{Conclusion and Future Work}

This study introduces CaseGPT, a novel framework for case-based reasoning that synergizes Large Language Models (LLMs) with Retrieval-Augmented Generation (RAG) technology. Our comprehensive experiments in medical and legal domains demonstrate CaseGPT's superior performance in case retrieval and insight generation, surpassing existing state-of-the-art methods.

\subsection{Key Contributions}

The primary contributions of this work can be summarized as follows:

\begin{description}
    \item[Innovative Architecture] CaseGPT seamlessly integrates semantic query understanding with precise case retrieval and insightful recommendations, advancing the field of AI-assisted professional decision-making.
    
    \item[Cross-Domain Efficacy] Empirical evidence shows significant improvements in retrieval accuracy and efficiency across multiple professional domains, highlighting the framework's versatility.
    
    \item[Expert-Validated Insights] The system generates high-quality, relevant insights based on retrieved cases, as corroborated by domain experts, bridging the gap between data retrieval and actionable knowledge.
    
    \item[Ethical Considerations] A nuanced discussion of the implications, limitations, and ethical considerations provides a balanced view of deploying such systems in sensitive professional contexts.
\end{description}

\subsection{Future Research Directions}

While CaseGPT marks a significant advancement, several critical areas warrant further investigation:

\begin{enumerate}
    \item \emph{Transfer Learning:} Exploring techniques to adapt CaseGPT to new professional domains with minimal retraining, enhancing its scalability and applicability.
    
    \item \emph{Explainable AI:} Developing advanced methods to elucidate the reasoning behind CaseGPT's recommendations, crucial for building trust and accountability in professional settings.
    
    \item \emph{Robustness Enhancement:} Investigating strategies to improve CaseGPT's performance on rare or unprecedented cases, ensuring reliability across diverse scenarios.
    
    \item \emph{Longitudinal Impact Assessment:} Conducting long-term studies to evaluate CaseGPT's influence on professional decision-making processes and outcomes, providing insights into its real-world efficacy.
    
    \item \emph{Privacy-Preserving Collaborative Learning:} Integrating federated learning approaches to facilitate knowledge sharing while maintaining data privacy, addressing a key concern in sensitive domains.
\end{enumerate}

\subsection{Concluding Remarks}

CaseGPT represents a paradigm shift in AI-assisted case-based reasoning for professional domains. By amalgamating advanced language understanding with robust retrieval and generation capabilities, it offers transformative potential in how professionals access, analyze, and leverage case data.

As we advance this technology, our focus must extend beyond technical refinements to encompass responsible development and thoughtful integration into professional practices. The true promise of CaseGPT lies in its capacity to augment human expertise, fostering improved outcomes across the critical domains of healthcare and law.

Ultimately, the success of CaseGPT will be measured not just by its technical prowess, but by its ability to:

\begin{itemize}
    \item Enhance decision-making processes in complex professional scenarios
    \item Democratize access to specialized knowledge and insights
    \item Uphold the highest standards of ethical practice and data privacy
\end{itemize}

As we look to the future, the continued development of CaseGPT must be guided by a commitment to these principles, ensuring that this powerful tool serves to elevate and support human expertise in the most critical and sensitive domains of professional practice.
\bibliographystyle{unsrt}

\end{document}